\documentclass[conference]{IEEEtran}
\IEEEoverridecommandlockouts
\usepackage{cite}
\usepackage{amsmath,amssymb,amsfonts,bm,amsthm,}
\usepackage{algorithmic}
\usepackage{graphicx}
\usepackage{textcomp}
\usepackage{xcolor}
\newtheorem{defi}{Definition}
\newtheorem{prop}{Proposition}

\newtheorem{thm}{Theorem}

\usepackage{subfigure}

\def\BibTeX{{\rm B\kern-.05em{\sc i\kern-.025em b}\kern-.08em
  T\kern-.1667em\lower.7ex\hbox{E}\kern-.125emX}}
\begin{document}

	\bibliographystyle{IEEEtran}
\title{Strategic Storage Investment in  Electricity Markets\\


\author{\IEEEauthorblockN{Dongwei Zhao\IEEEauthorrefmark{1}, Mehdi Jafari\IEEEauthorrefmark{2}, Audun Botterud\IEEEauthorrefmark{2}, Apurba Sakti\IEEEauthorrefmark{1}}
	\IEEEauthorblockA{\IEEEauthorrefmark{1}MIT Energy Initiative, Massachusetts Institute of Technology, Cambridge, USA\\\IEEEauthorrefmark{2}Laboratory for Information and Decision Systems (LIDS), Massachusetts Institute of Technology, Cambridge, USA}
	\vspace{-5ex}
}

}

%



\maketitle

\begin{abstract} Arbitrage is one important revenue source for energy storage in electricity markets. However, a large amount of storage in the market will impact the energy price and reduce potential revenues. This can lead to strategic behaviors of profit-seeking storage investors. To study the investors' strategic storage investments, we formulate a non-cooperative game between competing investors. Each investor decides the storage investment over a long investment horizon, and operates the storage for arbitrage revenues in the daily electricity market. Different investors can deploy storage with different characteristics. Their decisions are coupled due to the market price that is determined by all the investors' decisions. We use market data from California ISO to characterize the storage impact on the market price, based on which we establish a centralized optimization problem to compute the market equilibrium. We show that an increasing number of investors will increase the market competition, which reduces investors' profits but increases the total invested storage capacity. Furthermore,  we find that a slight increase in the storage efficiency (e.g., increased charge and discharge efficiency) can significantly improve an investor's profit share in the market.
\end{abstract}

\vspace{-2ex}

\section{Introduction}
\subsection{Background and motivation}

Energy storage can provide various applications (e.g., load shifting, frequency regulation, generation backup, transmission support) to the power grid and generate revenues to the investors\cite{overview1}. One of the most important applications and revenue streams is arbitrage in electricity markets. Storage operators can purchase energy and charge into storage when the price is low, and discharge when the price is high. According to the  US Energy Information Administration \cite{us2020battery}, arbitrage ranks top three among all the eleven storage applications in the US electricity  markets in 2019. In the future, the decreasing storage cost and active policy support are also expected to promote more storage deployment for arbitrage revenues \cite{sakti2018review}.

 However, the operation of large amounts of storage in the market will impact the market price. Studies show that the storage operation for arbitrage can reduce the price spread across time, and thus reduce the potential arbitrage revenues for investors \cite{shafiee2016economic}\cite{cui2017bilevel}. Such decreasing revenues may discourage profit-seeking investors from investing in storage.  {When there are multiple independent  investors in the  market, they will  compete for the limited revenues.}
Different investors can invest in storage with diverse characteristics, e.g., different technologies, which will affect their profit share in  competition.

The above discussions motivate us to answer the following key questions in this paper: 

\begin{itemize}
	\item \textit{How will the market equilibrium change with an increasing number of storage investors for arbitrage revenues?}
	\item \textit{What are the market positions of  investors who have storage technologies with heterogeneous characteristics?}
\end{itemize}

Our work focuses on the example of energy  arbitrage  in studying competition impact. We will extend the competition model to multiple storage applications in future studies.

\vspace{-0.5ex}
\subsection{Related work}
\vspace{-0.5ex}

There have been a series of active studies on the arbitrage revenues of energy storage. One class focuses on pricing-taking storage (e.g., \cite{Wang2018storage,Xu2020storageprice,shapiro2020real}), which ignores the storage impact on the market price. A second class considers storage as price makers, and our work also lies in this class.
For price-making storage, some literature (e.g., \cite{shafiee2016economic,cui2017bilevel, cruise2018impact}) only consider the storage operation and ignore the decision and cost of storage investment. This fails to fully evaluate the storage values since the high investment cost is vital in the decision making. Other literature consider the storage investment, but oftentimes assume only one investor or multiple cooperative investors (e.g., \cite{nasrolahpour2016strategic,Xu2017storagestack,Huang2019storage}). In practice, the storage investors are usually independent and compete in the market.

The work that is most closely related to ours is \cite{Qin2019storage}, in which Qin \textit{et al.} studied strategic storage investment among non-cooperative investors. Our work differs from \cite{Qin2019storage} in several crucial ways. First, Qin \textit{et al.} focused on aggregated end-user investors, which modeled those small investors as a
continuum. This setting cannot be applied to large investors in the wholesale market. In contrast, our work models large investors and explores the impact of the number of investors. Second, the model in \cite{Qin2019storage} only considered homogeneous storage among all the investors, which ignored the power rating and duration decisions. In contrast, our work models heterogeneous-storage investors, while also considering power rating decisions and duration constraints.

\vspace{-0.5ex}
\subsection{Main results and contributions}
\vspace{-0.5ex}
 Our work studies the strategic storage investment of multiple investors. Those investors can invest in heterogeneous storage and compete for arbitrage revenues. At the beginning of an investment horizon (e.g., 10 years), each investor decides the investment of storage. Next, 
 for each day in the investment horizon, each investor determines the charge and discharge profiles over times slots for seeking arbitrage revenues. Investors' storage decisions are coupled due to the market price that is determined by all the investors' storage operation, so we model the storage-investment problem as a non-cooperative game. We use the market data from CAISO to characterize the storage impact on the market price, based on which we establish a centralized optimization problem to compute the Nash equilibrium. Based on the CAISO data, we demonstrate  market equilibrium results with competing investors. We show that a higher number of investors increases the market competition, while an improvement in one investor's storage efficiency has a large impact on his investment decision.

The main contributions of this paper are listed as follows.

\begin{itemize}
\item \textit{Strategic storage investment:} To the best of our knowledge, our work is the first to study the strategic storage investment of heterogeneous investors who compete for arbitrage revenues. Our model provides insights into  the equilibrium of market competition with an increasing number of heterogeneous storage investors.

\item \textit{Computing Nash equilibrium:} Considering the storage impact on the market price, we model the storage-investment problem as a non-cooperative game. We use the market data from CAISO to characterize the storage impact on the market price, based on which we prove the existence of a pure strategy Nash equilibrium. We establish a centralized convex optimization problem to compute the equilibrium, which can be efficiently solved.

\item \textit{Practical insights:} Based on the CAISO  market data, we show that an increasing number of storage investors increases the market competition. This reduces storage investors' profits but increases the total invested capacity. Furthermore, somewhat surprisingly, we find that a slight increase in the storage efficiency (e.g., increased charge and discharge efficiency) can substantially increase the profit share of an investor in the market.
\end{itemize}

\vspace{-0.5ex}
\section{System model}
\vspace{-0.5ex}

We consider a set of non-cooperative storage investors $\mathcal{I}=\{1,\dots,I\}$.
They compete in investing energy storage to get arbitrage revenues in the electricity market. Next, we will  introduce the timescale of investors' decision-making.

\vspace{-0.5ex}
\subsection{Timescale of decision-making}
\begin{figure}[t]
	\centering
	\includegraphics[width=2.4in]{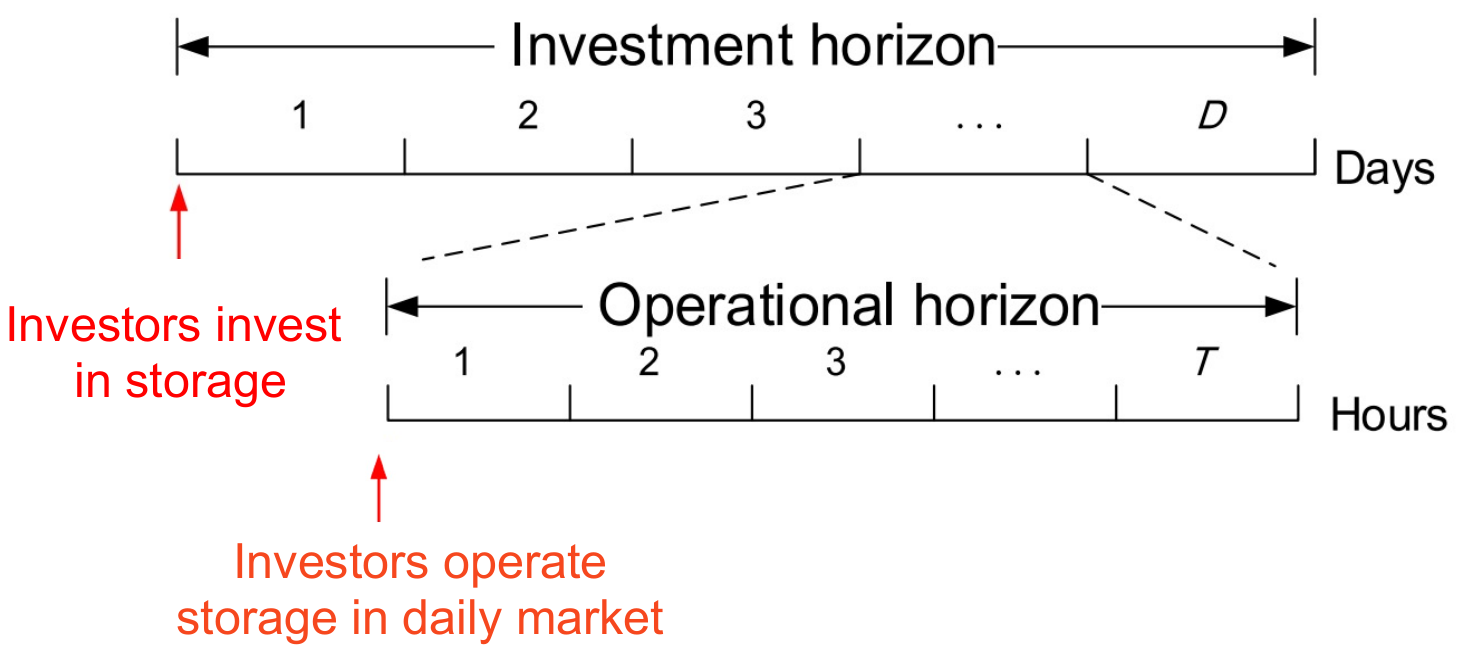}
	\vspace{-1.5ex}
	\caption{Timescales.}
	\label{fig:time0}
	\vspace{-3ex}
\end{figure}

\vspace{-0.5ex}
Figure \ref{fig:time0} illustrates two timescales of decision making in our model. At the beginning of an investment horizon $\mathcal{D}\hspace{-1mm}=\hspace{-1mm}\{1,2,...,D\}$ of $D$ days (e.g., $D$ corresponding to many years), each investor $i\in \mathcal{I}$ decides the energy capacity $S_i$ and power rating $P_i$ of the storage.

The investment horizon is divided into many operational horizons, i.e., each day $d\in \mathcal{D}$ corresponds to one operational horizon, which is further divided into multiple time slots $\mathcal{T}\hspace{-1mm}=\hspace{-1mm}\{1,2,...,T\}$ (e.g., 24 time slots corresponding to 24 hours). For each day $d \in \mathcal{D}$, each investor $i$ decides the charge and discharge profiles over times slots $\mathcal{T}$, i.e., $\bm{p}_i^{ch,d}=({p}_i^{ch,d}[t], t\in \mathcal{T})$ and $\bm{p}_i^{dis,d}=({p}_i^{dis,d}[t], t\in \mathcal{T})$. We also denote the energy level in investor $i$'s storage by $\bm{e}_i^d=\{e_i^d[t],~ \forall t \in \mathcal{T}'\}$ for day $d \in \mathcal{D}$, where $\mathcal{T}'=\{0\}\bigcup\mathcal{T}$ and $e_i^d[0]$ denotes the initial energy level. We denote by a vector $\bm{x}_i=\Big(S_i, P_i, ( \bm{p}_i^{ch,d}, d\in  \mathcal{D}),$ $ (\bm{p}_i^{dis,d}, d\in  \mathcal{D}), ( \bm{e}_i^{d}, d\in \mathcal{D})\Big)$ all the decision variables of investor $i$ over the two timescales. 

To facilitate the formulation, we will calculate the  investors' expected profits based on the timescale of one day. We scale investors' investment costs into one day and calculate the daily expected arbitrage revenues. We use historical data to construct a scenario set $\Omega$ that represents all the days in the investment horizon, where one day lies in the scenario $\omega \in \Omega$ with a probability $\rho^\omega$. We denote the original market price without storage operation by $\pi_0^\omega[t],~\forall t\in\mathcal{T}$. For example, if we consider 365 scenarios (days)  based on one-year market-price data,  we let each scenario have the probability of 1/365 and calculate the expected arbitrage revenues based on the probabilities of 365 scenarios. To further reduce the computation burden, in the numerical study, we construct 12 scenarios corresponding to 12 months, which we will explain more in Section \ref{sec:sim}.A. We rewrite by $\bm{x}_i=\Big(S_i, P_i, ( \bm{p}_i^{ch,\omega}, \omega\in  \Omega),$ $ (\bm{p}_i^{dis, \omega}, \omega\in  \Omega), ( \bm{e}_i^{\omega}, \omega\in  \Omega)\Big)$  the decision vector of each investor $i$ over the two timescales. Next, we introduce investors' profits, storage constraints, and game-theoretic model. 

\vspace{-0.2ex}
\subsection{Profit of investors}
\vspace{-0.2ex}
The profit of each investor includes the arbitrage revenues, deducting the storage investment and operation cost. 

For the arbitrage revenues, in scenario $\omega$, we denote the market price with storage impact by $\bm{\pi}^\omega=({\pi}^\omega[t],t\in \mathcal{T})$. Note that the market price is affected by all the users' storage decisions. Thus, we denote the market price of time slot $t$ in scenario $\omega$ as a function of all the investors' decision variables, i.e.,  ${\pi}^\omega[t](\bm{x}_1, \bm{x}_2\ldots \bm{x}_I)$. We will introduce the model of the market price in Section \ref{section:price} later. The expected revenues of investor $i$ on a daily basis is  
\vspace{-0.5ex}
\begin{align}
&~~~~R_i(\bm{x}_i, \bm{x}_{-i})\notag\\
&=\mathbb{E}_{\omega \in \Omega} \Big[\sum_{t\in \mathcal{T}} (-{p}_i^{ch,\omega}[t]+{p}_i^{dis,\omega}[t]) \cdot \pi^\omega[t] (\bm{x}_i, \bm{x}_{-i})\Big],
\end{align}
where $\bm{x}_{-i}$ denotes all the other investors' decision variables other than investor $i$.

The storage investment cost of investor $i$ includes the capital costs for both energy capacity and power rating, which we characterize in  $C^{inv}(\bm{x}_i)$.
\vspace{-0.5ex}
\begin{align}
C_i^{inv}(\bm{x}_i)=\kappa c_i^{S}S_i+\kappa c_i^{P}P_i,
\end{align}
where $c_i^{S}$ and $c_i^{P}$ denote the unit costs for capacity and power rating, respectively. The coefficient $\kappa$ scales the capital cost into one day,  the details of which  can be found in \cite{zhao2019virtual}.


We also consider the storage operation cost incurred by charge and discharge, which can include the degradation cost and maintenance cost. We adopt a linear model for the operation cost and calculate the expected operation cost of investor $i$ on a daily basis in the following.
\vspace{-0.5ex}
\begin{align}
C_i^{opr}(\bm{x}_i)=\mathbb{E}_{\omega \in \Omega} \Big[\sum_{t\in \mathcal{T}}\left(c_i^{ch}{p}_i^{ch,\omega}[t]+c_i^{dis}{p}_i^{dis,\omega}[t]\right)\Big],
\end{align}
where $c_i^{ch}$ and $c_i^{dis}$ denote the unit costs for charge and discharge amount, respectively. 

Overall, we have the profit $f_i(\bm{x}_i, \bm{x}_{-i})$ of investor $i$ over the investment horizon (scaled in one day) as follows.
		\begin{align}
	f_i(\bm{x}_i, \bm{x}_{-i})=R_i(\bm{x}_i, \bm{x}_{-i})-C_i^{inv}(\bm{x}_i)-C_i^{opr}(\bm{x}_i).
\end{align}
Note that investor $i$'s profit is coupled with others due to the market price that is determined by all the investors‘ decisions. 

\vspace{-0.5ex}
\subsection{Decision set of investor $i$}

 We introduce the constraints for the storage investment and operation in the following.
		\begin{align}
&0\leq 	p_i^{dis,\omega}[t]\leq P_i,\forall t\in\mathcal{T},\forall \omega\in\Omega, \label{eq:ch}\\
	&0\leq 	p_i^{ch,\omega}[t]\leq P_i,\forall t\in\mathcal{T},\forall \omega\in\Omega, \label{eq:dis}\\
& e_i^\omega[t]\hspace{-0.3mm}=\hspace{-0.3mm}e_i^\omega[t-1]\hspace{-0.3mm}+\hspace{-0.3mm}\eta_i^cp_i^{ch,\omega}[t]\hspace{-0.3mm}-\hspace{-0.3mm}p_i^{dis,\omega}[t]/\eta_i^d, \forall t\in\mathcal{T},\forall \omega\in\Omega, \label{eq:dynamics1}\\
	&0\leq e_i^\omega[t]\leq S_i,\forall t\in\mathcal{T}',\forall \omega\in\Omega, \label{eq:dynamics2}\\
	&e_i^\omega[0]= e_i^\omega[T],\forall \omega\in\Omega,\label{eq:dynamics3}\\
&\underline{\alpha}_i\leq \frac{S_i}{P_i}\leq \overline{\alpha}_i. \label{eq:duration}
\end{align}

We explain the constraints in detail. First, equations \eqref{eq:ch}-\eqref{eq:dis} are the power rating limits for the charge and discharge, respectively.\footnote{We do not enforce the constraints of non-simultaneous charge and discharge. In the simulation, simultaneous charge and discharge never happens.}
Second, equations \eqref{eq:dynamics1}-\eqref{eq:dynamics3} are the constraints for the energy level in the storage. Equation \eqref{eq:dynamics1} is the energy level change with time due to the charge and discharge operation, where $\eta_i^c$ and $\eta_i^d$ denote the charge and discharge efficiency, respectively. Equation \eqref{eq:dynamics2} shows the energy level is constrained by the storage capacity. Equation \eqref{eq:dynamics3} makes the initial energy level equal to the final level for each day, which decouples the storage operation across different days. Finally, equation \eqref{eq:duration} enforces a general duration limit on the invested capacity and power rating, where $\underline{\alpha}_i$ is the lower bound and $\overline{\alpha}_i$ is the upper bound. We denote by $\mathcal{X}_i$ investor $i$'s decision set that is established by \eqref{eq:ch}-\eqref{eq:duration}. 

\vspace{-0.5ex}
\subsection{Storage-investment game model}
 
Based on the profits and decision sets of the investors, we formulate the storage-investment game $G$ among investors.

\begin{itemize}
	\item \textit{Players}: Investor $i\in \mathcal{I}$
	\item \textit{Strategy}: $\bm{x}_i\in \mathcal{X}_i$ of investor $i$ 
	\item \textit{Profit}: $f_i(\bm{x}_i,\bm{x}_{-i})$ of investor $i$ 	
\end{itemize}

We assume that the investors make decisions simultaneously, and they do not observe other investors' decisions when making decisions.\footnote{Our work focuses on the one-time investment decisions of all the investors, which does not consider multi-stage sequential decisions. The impact of investors who enter the market earlier and later can be further explored in the future studies.} However, each investor knows the information of market price function and others' storage parameters.

\section{Market-price function}\label{section:price}

The storage investors' decisions are coupled due to the market price $\pi^\omega (\bm{x})$. 
We will first introduce a linear model for the price based on the aggregate storage operation of all the investors. Then, we use the data from CAISO day-ahead markets to characterize such a linear price function.

\subsection{Model of market price function}

We assume that the marked price is linear with respect to the net charge and discharge amount of storage in the market, which is shown as follows.
\vspace{-0.5ex}
\begin{align}
 \pi^\omega [t](\bm{x}_1\ldots \bm{x}_I)\hspace{-0.45mm}=\hspace{-0.45mm} \pi^\omega_0 [t]\hspace{-0.4mm}-\hspace{-0.4mm}{a^\omega [t]}\hspace{-0.45mm}\cdot \hspace{-0.45mm}p^\omega[t], \forall t \in \mathcal{T},\forall \omega\in \Omega, \label{eq:price}
\end{align}
where we let $p^\omega[t]=\sum_{i\in \mathcal{I}} (-{p}_i^{ch,\omega}[t]+{p}_i^{dis,\omega}[t])$. The price $\pi^\omega_0 [t]$ is the original market price without storage operation. The coefficient $a^\omega[t]$ is the slope  that we will characterize later. The part  $p^\omega[t]$ is the net discharge amount of all the investors in the market and we explain it next.

 We examine the storage impact on the market price analogous to renewable energy. In the current practice, the market price is usually set at the system marginal cost\cite{fundamentals}. The addition of zero-marginal-cost renewable energy in the system can be treated as a decrease in the net demand for the high-marginal-cost energy sources, which reduces the market price. This 
depressed price due to the increasing renewable energy has been demonstrated in some electricity markets, such as 
California and Western Australia \cite{leslie2020designing}. The storage also usually has a low marginal operation cost \cite{Xu2017storagestack} \cite{leslie2020designing}, which can also be charged with the cheapest resources in the system. Thus, we assume that the storage has a similar impact on the market price like renewable energy. As shown in \eqref{eq:price}, on one hand, if $p^\omega[t]>0$, it means that the aggregate storage operation of all the investors leads to net discharge amount in the market. This reduces the net demand and thus reduces the market price. On the other hand, the negative $p^\omega[t]<0$ means the net charge amount in the market, which increases the net demand and thus increases the market price.

\subsection{Characterization of market price function}

We use the market data in CAISO to characterize the market price function \eqref{eq:price}. We examine one-year data (from January 2019  to December 2019) of the day-ahead (uncongested energy) prices, and day-ahead forecast net demand (system total demand deducting system total renewable energy)\cite{calidata}.
\begin{figure}[t]
	\centering
\includegraphics[width=3in]{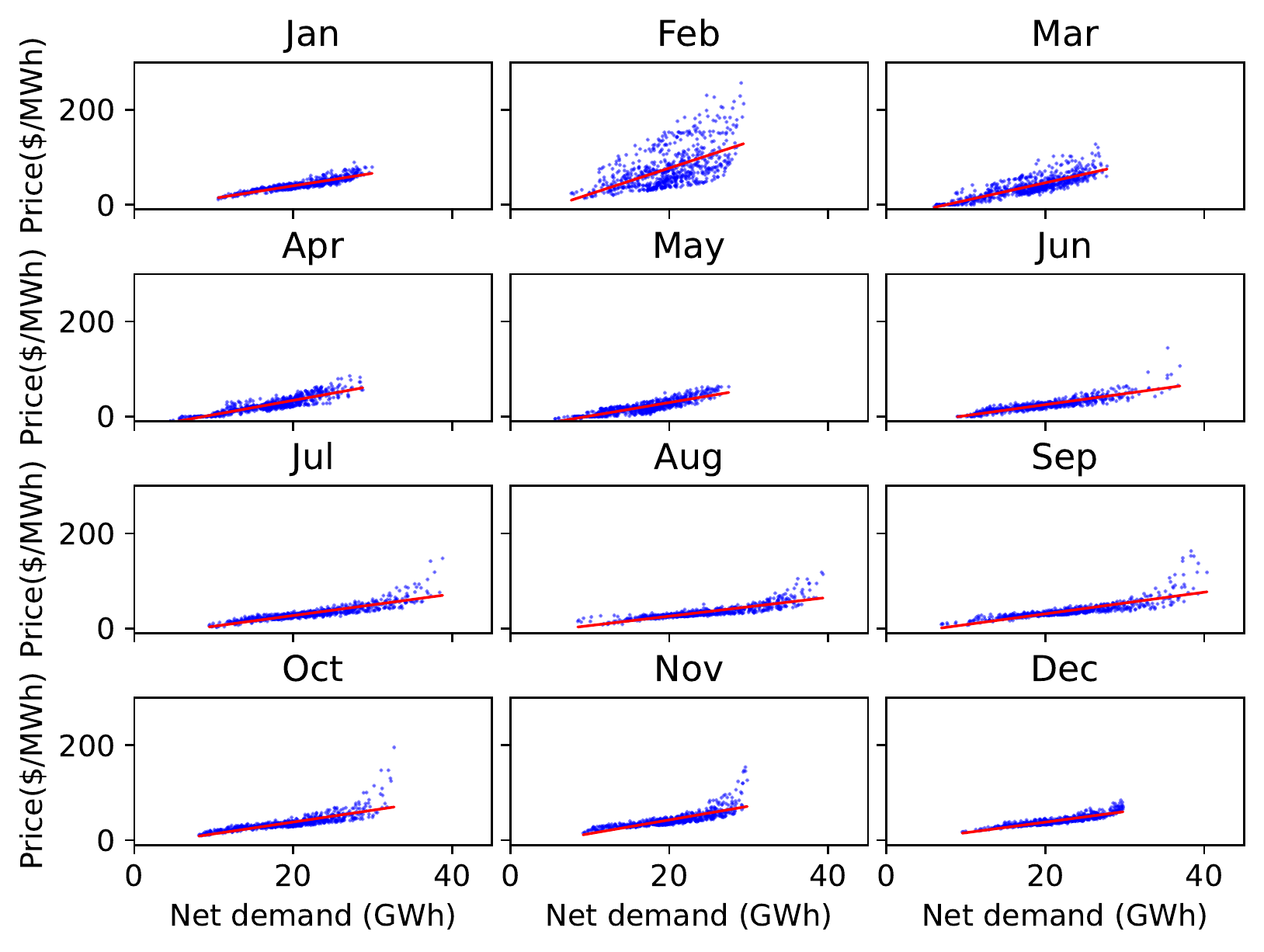}
\vspace{-2ex}
	\caption{Day-ahead price with forecast net demand in 12 months of 2019.}
	\label{fig:caiso_price_data}
	\vspace{-2ex}
\end{figure}

In Figure \ref{fig:caiso_price_data}, we plot the data of day-ahead prices and net demand in blue points, for all the hours in each month from January to December.
Figure \ref{fig:caiso_price_data} shows that the linear model provides a good approximation for the relationship between the price and net demand. We use the linear regression to characterize the linear approximation functions.\footnote{ The market-price function in our model actually approximates the supply curve of conventional generators in the market. A piece-wise linear structure  or modeling based on actual dispatch curves can be more accurate. However, it also significantly increases the complexity of analysis and computation for Nash equilibrium, which we leave as future work.} We set the slope of the linear approximation function of each month as a uniform slope $a$ for all the hours in this month. Next, we will show how to compute the Nash equilibrium based on such a market price assumption.

\vspace{-1ex}
\section{Solution method}
\vspace{-1ex}

Under the linear price function \eqref{eq:price}, we show that the pure strategy Nash equilibrium of the storage-investment game $G$ exists. We establish an equivalent centralized optimization problem to compute the equilibrium.

We first introduce the definition of the pure strategy Nash equilibrium for the storage-investment game $G$.

\vspace{-0.5ex}
\begin{defi}[Pure strategy Nash equilibrium]\label{def:equilibrium}
In the storage-investment game $G =\langle \mathcal{I},(\mathcal{X}_i),(f_i)\rangle$, the strategy profile $\bm{x}^*\in \Pi_{i\in \mathcal{I}} \mathcal{X}_i$ is called a pure strategy Nash equilibrium if for every investor $i\in \mathcal{I}$, $f_i(\bm{x}_i^*,\bm{x}_{-i}^*)\geq f_i(\bm{x}_i,\bm{x}_{-i}^*)$ for any $\bm{x}_i\in \mathcal{X}_i$.
\end{defi}
\vspace{-0.5ex}

\noindent The above definition states that the pure strategy Nash equilibrium is a state where every investor's decision maximizes his profit given other investors' decisions.

 We can prove that given the price function \eqref{eq:price}, the pure strategy Nash equilibrium of the game $G$ always exists, which is shown in Proposition \ref{prop:existence}. 

\vspace{-0.5ex}
\begin{prop}[Existence of Nash equilibrium] \label{prop:existence}
	The storage-investment game $G$ with the market price function \eqref{eq:price} exists a pure strategy Nash equilibrium. 
\end{prop}
\vspace{-0.5ex}

 We can further establish an equivalent centralized optimization problem to compute the Nash equilibrium in Theorem \ref{thm:compute}.

\vspace{-0.5ex}
\begin{thm}[Computing Nash equilibrium]\label{thm:compute}
	One Nash equilibrium of the game $G$ is equivalent to the solution to the following convex optimization problem, where we let $A^\omega(\bm{x}_i,\bm{x}_j):=(-{p}_i^{ch,\omega}[t]+{p}_i^{dis,\omega}[t])\cdot (-{p}_j^{ch,\omega}[t]+{p}_j^{dis,\omega}[t])$.
	\begin{align}
		\max_{\bm{x}}~\hspace{-1.2mm}	&\sum_{i\in \mathcal{I}}\hspace{-1mm} f_i(\bm{x})\hspace{-0.5mm}+\hspace{-0.5mm}\mathbb{E}_{\omega \in \Omega}\Big[\sum_{t\in \mathcal{T}}\sum_{1\leq i<j\leq I}\hspace{-1.2mm} a^\omega[t]\hspace{-0.5mm}\cdot \hspace{-0.5mm}A^\omega(\bm{x}_i,\bm{x}_j)\Big],\label{eq:nash}\\
		\text{s.t.}~	&\bm{x}_i\in \mathcal{X}_i,\forall i\in \mathcal{I}.\notag
	\end{align}
\end{thm}
\vspace{-0.5ex}

We prove Proposition \ref{prop:existence} and Theorem \ref{thm:compute}  by showing the KKT conditions of the centralized  optimization problem in Theorem \ref{thm:compute}  are equivalent to those for the Nash equilibrium.  We omit the detailed proof due to page limit.
In the objective \eqref{eq:nash}, the first part $\sum_{i\in \mathcal{I}} f_i(\bm{x})$ is the total profit of all the investors. The second part with $A^\omega(\bm{x}_i,\bm{x}_j)$ is an additional item brought by the market competition. 
Furthermore, the optimization problem in Theorem \ref{thm:compute} is quadratic programming, which can be efficiently solved. Next, we conduct simulations to show the market equilibrium results with competing investors.

\section{Numerical study} \label{sec:sim}

We use the market data from CAISO to perform the simulation\cite{calidata}. We show how the market equilibrium will change with the increasing number of investors and show the impact of  storage heterogeneity on the investor's market share.

\subsection{Simulation setup}

We construct 12 scenarios corresponding to 12 months from  CAISO one-year data in 2019. Each scenario has a probability of 1/12. More specifically, for the slope $a^\omega$ of the market-price function, we characterize a uniform slope for all the hours in each month as shown in Figure \ref{fig:caiso_price_data}. For the original market prices $\pi_0^\omega$, we select a representative daily price profile for each month. The criterion is to select the profile that is closest to the average price profile of all days in that month.

To demonstrate the market competition equilibrium, in the simulation we set a low capital cost for the storage: energy-capacity cost 90\$/kWh and power-rating cost 180\$/kW, with lifespan 20 years and annual interest rate= $5\%$.

\subsection{Equilibrium with an increasing number of storage investors}

We show that an increasing number of investors increases the market competition, which reduces storage investors' profits but increases the total invested capacity. 

\begin{figure}[t]
	\centering
	\hspace{-3ex}
	\subfigure[]{
		\raisebox{-2mm}{\includegraphics[width=1.7in]{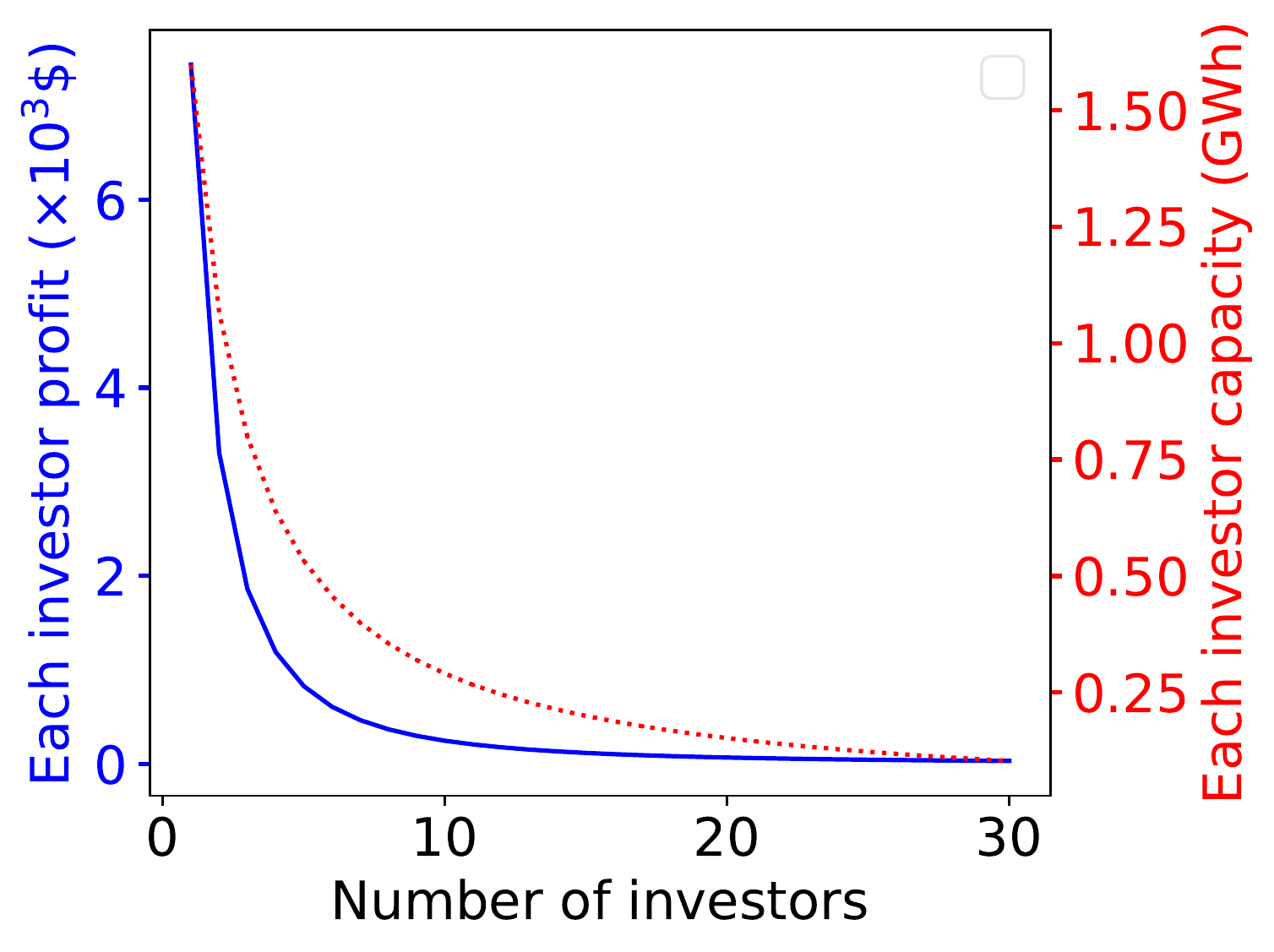}}}
	\hspace{-1.5ex}
	\subfigure[]{
		\raisebox{-2mm}{\includegraphics[width=1.7in]{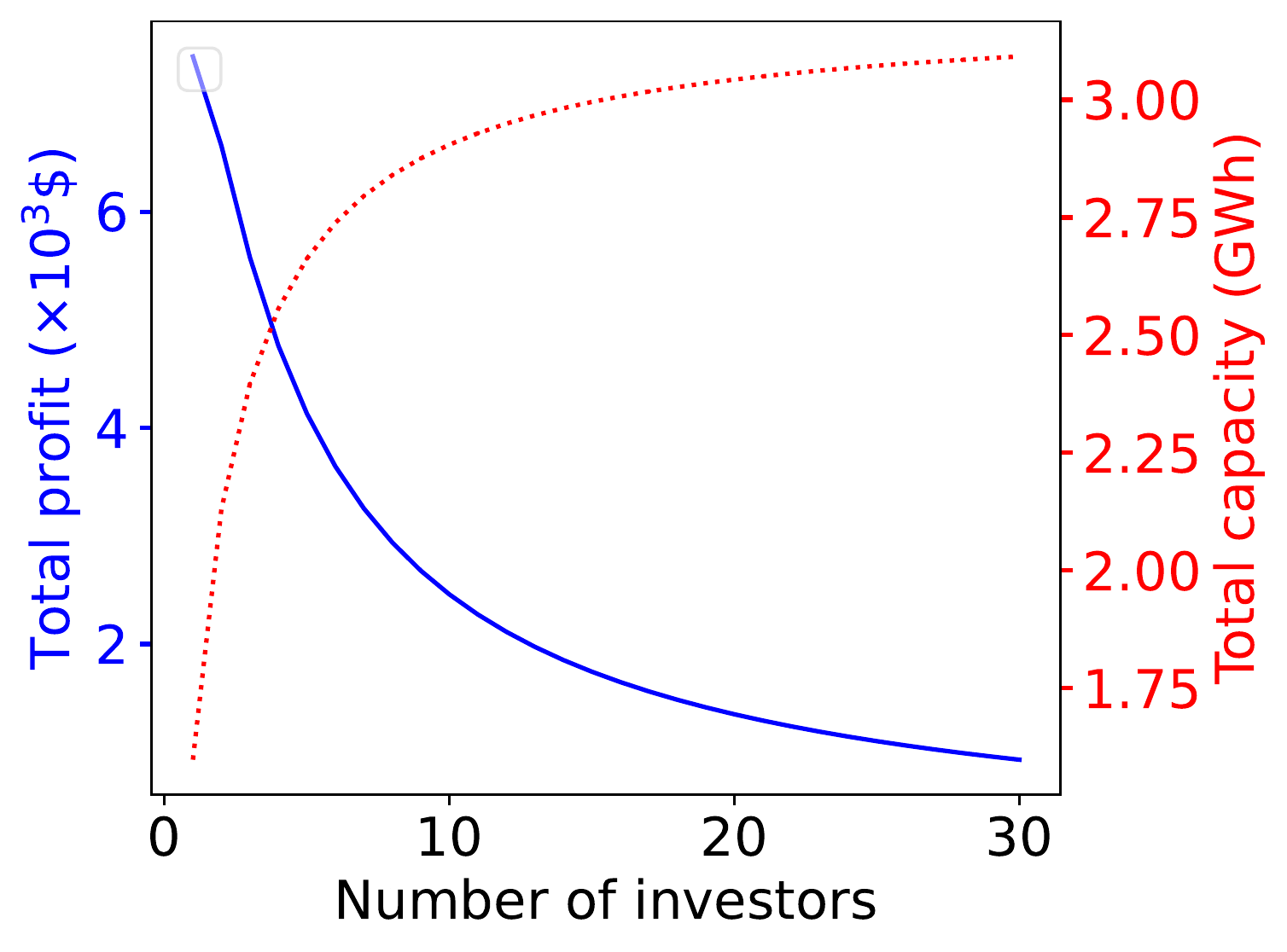}}}
	\vspace{-2mm}
	\caption{(a) \small Each investor's profit and invested storage capacity; (b) Total profit and capacity. Both are with the number of investors.}
	\label{fig:num}
	\vspace{-5mm}
\end{figure}

To clearly show the impact of the number of investors, we consider investors with homogeneous storage. In Figure \ref{fig:num}, we vary the number of investors in the market and show how users' profits and invested storage capacities change accordingly at the equilibrium. In Figure \ref{fig:num}(a), we show each investor's profit (in blue) and invested storage capacity (in red). In Figure \ref{fig:num}(b), we show all the investors' total profit (in blue) and total capacity (in red). We have the following observation.

 \vspace{0.4ex}

\noindent \textbf{Observation 1}:  \textit{An increasing number of investors will reduce investors' profits but increase the total invested capacity.}
\vspace{0.4ex}

We can see that in Figure \ref{fig:num}(a) and (b), each investor's profit and all the investors' total profit will both decrease with the number of investors, which is due to the increased competition in the market. However, although each investor's invested capacity decreases shown in Figure \ref{fig:num}(a), the total capacity increases as in Figure \ref{fig:num}(b). The increased capacity implies a benefit to the system by shaving the peak demand.

\vspace{-0.2ex}
\subsection{Market positions of heterogeneous investors}
\vspace{-0.2ex}

We investigate the market equilibrium with heterogeneous investors, and
find that a slight increase in performance (e.g., increased charge and discharge efficiency) can significantly improve the profit share of an investor in the market.\footnote{We demonstrate one example of charge and discharge efficiency due to page limit. We will discuss more parameters. e.g., capital costs and duration in the coming extended version.}

\begin{figure}[t]
	\centering
	\hspace{-3ex}
	\subfigure[]{
		\raisebox{-2mm}{\includegraphics[width=1.7in]{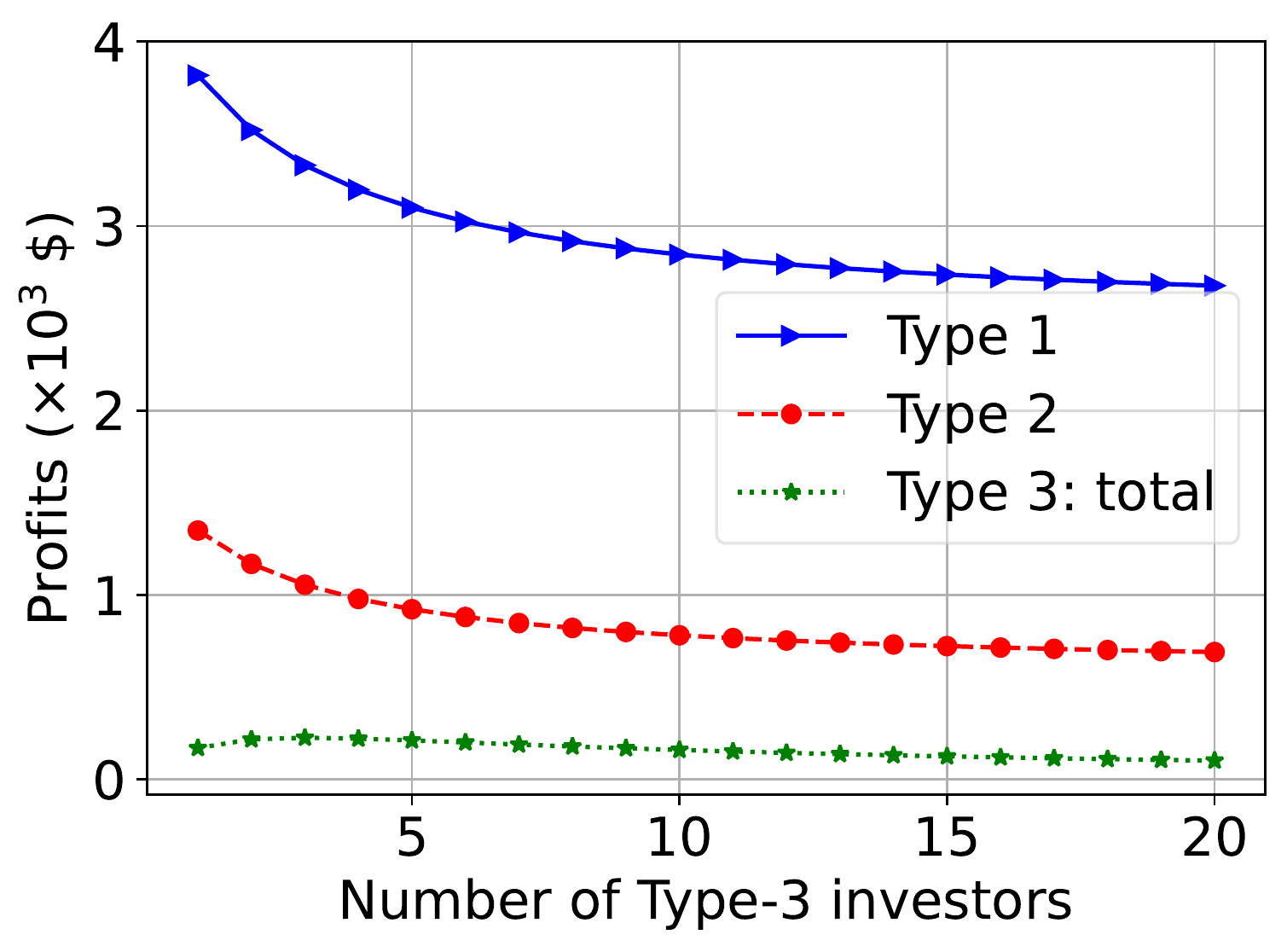}}}
	\hspace{-2ex}
	\subfigure[]{
		\raisebox{-2mm}{\includegraphics[width=1.7in]{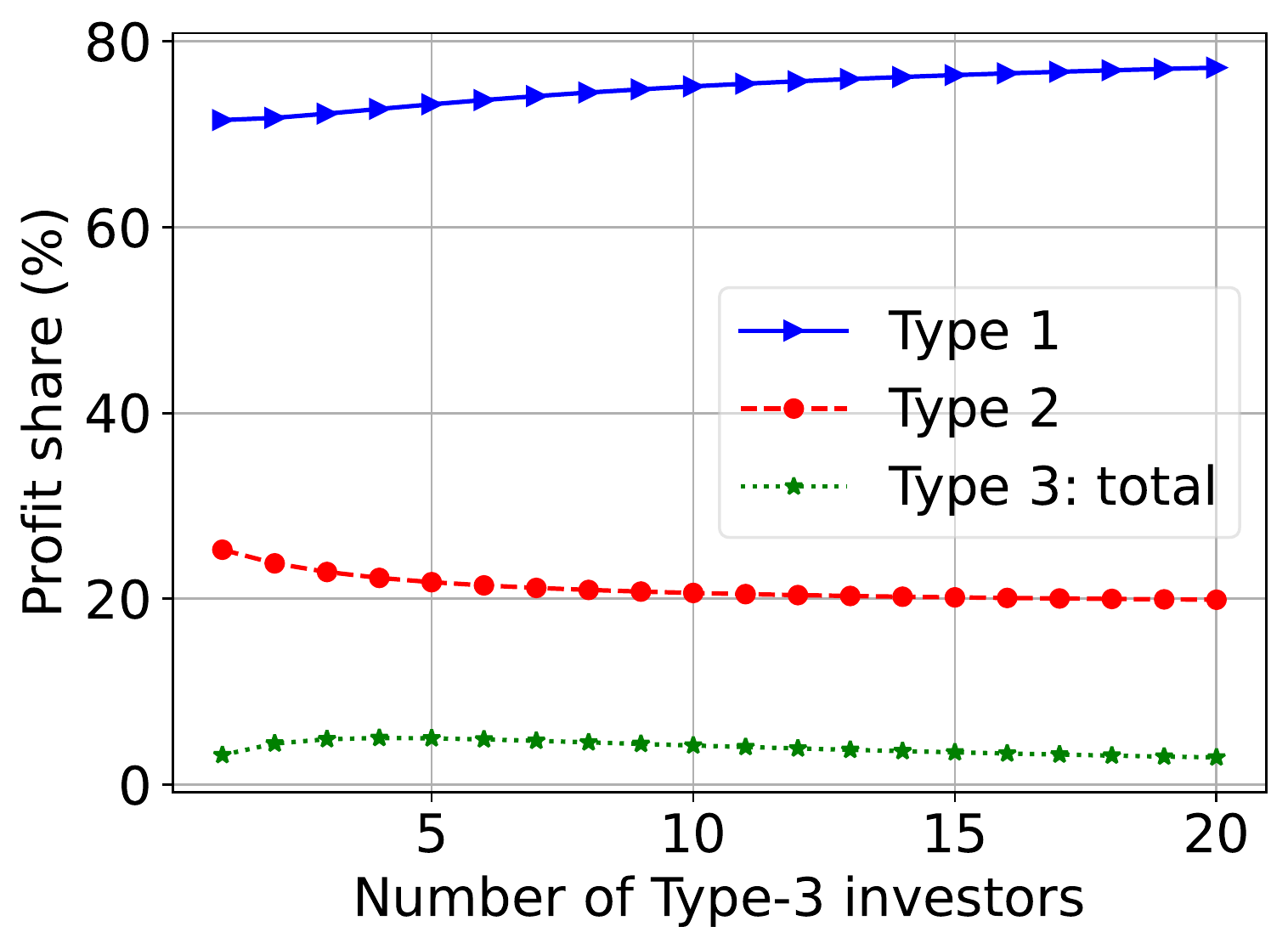}}}
	\vspace{-2mm}
	\caption{(a) \small Profits; (b) Profit share. With number of Type-3 investors.}
	\label{fig:heter}
	\vspace{-5mm}
\end{figure}

 We consider three types of investors whose storage only differs in the charge and discharge efficiency. We assume that Type 1, Type 2, and Type 3 have the charge and discharge efficiency at $\eta^c=\eta^d=0.95,~0.94,$ and $0.93$, respectively. For Type 1 and Type 2, we consider only one investor for each type. For Type 3, we vary the number of its investors in the market from 1 to 20, and in Figure \ref{fig:heter} we show how the equilibrium profit of each type will change accordingly.
 
 In Figure \ref{fig:heter}(a), we show the profits of Type 1 (blue curve) and Type 2 (red curve), and the total profit of all Type-3 investors (green curve).
 In Figure \ref{fig:heter}(b), we show the profit share percentage of Type 1 (blue curve), Type 2 (red curve), and all the investors of Type 3 (green curve) in the market. We have the following observations.

 \vspace{0.4ex}
 
 \noindent \textbf{Observation 2}: \textit{A slight advantage of storage efficiency makes the investor dominate the market.}
  \vspace{0.4ex}
  
  As we can see in Figure \ref{fig:heter}(a), Type-1 investor has a much higher profit than Type 2 and Type 3 due to a slightly higher charge and discharge efficiency. The total profits of all Type-3 users are even smaller than the single Type-2 investor. Also, as shown in Figure \ref{fig:heter}(b),  the single Type-1 investor occupies 70\% of the profit in the market.

  \vspace{0.4ex}
 \noindent \textbf{Observation 3}: \textit{An increasing number of low-efficiency investors can even increase the profit share of the high-efficiency investors in the market.}
  \vspace{0.4ex}
  
In Figure \ref{fig:heter}(a), as the number of Type-3 investors increases, the profits of Type 1 and Type 2 decrease due to the reduced arbitrage revenues. However, the increasing number of Type 3 always increases the profit share of the most efficient Type-1 investor as in Figure \ref{fig:heter}(b). The reason is that the increasing number of low-efficiency investors significantly increases the competition between themselves. This has a limited impact on the high-efficiency ones and even increases their profit share. 

Observation 2 and Observation 3 also suggest that in order to be at advantage in the market, the investors should try to improve their storage efficiency, which can potentially facilitate the development of storage technology.

\section{Conclusion}
\vspace{-0.2ex}

This paper studies the strategic storage-investment problem. We formulate a game-theoretic model between competing investors who can have storage technologies with diverse characteristics. We use market data from CAISO to characterize the storage impact on the market price, based on which we compute the market equilibrium of the storage investment and operation by establishing a centralized optimization problem. Using the CAISO market data, we show that the increasing number of investors increases the market competition, which reduces storage investors' profits but increases the total invested capacity. Furthermore, somewhat surprisingly, we find that a slight increase in the storage efficiency has a large impact on the investor's profit share in the market. In future work, we will further explore the impact of storage duration and renewable energy on the market equilibrium.


\bibliography{storage}

\begin{thebibliography}{10}
\providecommand{\url}[1]{#1}
\csname url@samestyle\endcsname
\providecommand{\newblock}{\relax}
\providecommand{\bibinfo}[2]{#2}
\providecommand{\BIBentrySTDinterwordspacing}{\spaceskip=0pt\relax}
\providecommand{\BIBentryALTinterwordstretchfactor}{4}
\providecommand{\BIBentryALTinterwordspacing}{\spaceskip=\fontdimen2\font plus
\BIBentryALTinterwordstretchfactor\fontdimen3\font minus
  \fontdimen4\font\relax}
\providecommand{\BIBforeignlanguage}[2]{{%
\expandafter\ifx\csname l@#1\endcsname\relax
\typeout{** WARNING: IEEEtran.bst: No hyphenation pattern has been}%
\typeout{** loaded for the language `#1'. Using the pattern for}%
\typeout{** the default language instead.}%
\else
\language=\csname l@#1\endcsname
\fi
#2}}
\providecommand{\BIBdecl}{\relax}
\BIBdecl

\bibitem{overview1}
X.~Luo, J.~Wang, M.~Dooner, and J.~Clarke, ``Overview of current development in
  electrical energy storage technologies and the application potential in power
  system operation,'' \emph{Applied Energy}, vol. 137, pp. 511--536, 2015.

\bibitem{us2020battery}
U.~E. I.~A. (EIA), ``Battery storage in the united states: An update on market
  trends,'' 2020.

\bibitem{sakti2018review}
A.~Sakti, A.~Botterud, and F.~O’Sullivan, ``Review of wholesale markets and
  regulations for advanced energy storage services in the united states:
  Current status and path forward,'' \emph{Energy policy}, vol. 120, pp.
  569--579, 2018.

\bibitem{shafiee2016economic}
S.~Shafiee, P.~Zamani-Dehkordi, H.~Zareipour, and A.~M. Knight, ``Economic
  assessment of a price-maker energy storage facility in the alberta
  electricity market,'' \emph{Energy}, vol. 111, pp. 537--547, 2016.

\bibitem{cui2017bilevel}
H.~Cui, F.~Li, X.~Fang, H.~Chen, and H.~Wang, ``Bilevel arbitrage potential
  evaluation for grid-scale energy storage considering wind power and lmp
  smoothing effect,'' \emph{IEEE Transactions on Sustainable Energy}, vol.~9,
  no.~2, pp. 707--718, 2017.

\bibitem{Wang2018storage}
H.~Wang and B.~Zhang, ``Energy storage arbitrage in real-time markets via
  reinforcement learning,'' in \emph{2018 IEEE Power Energy Society General
  Meeting (PESGM)}, 2018, pp. 1--5.

\bibitem{Xu2020storageprice}
B.~Xu, M.~Korpås, and A.~Botterud, ``Operational valuation of energy storage
  under multi-stage price uncertainties,'' in \emph{2020 59th IEEE Conference
  on Decision and Control (CDC)}, 2020, pp. 55--60.

\bibitem{shapiro2020real}
C.~R. Shapiro, C.~Ji, and D.~F. Gayme, ``Real-time energy market arbitrage via
  aerodynamic energy storage in wind farms,'' in \emph{2020 IEEE ACC}, 2020,
  pp. 4830--4835.

\bibitem{cruise2018impact}
J.~R. Cruise, L.~Flatley, and S.~Zachary, ``Impact of storage competition on
  energy markets,'' \emph{European Journal of Operational Research}, vol. 269,
  no.~3, pp. 998--1012, 2018.

\bibitem{nasrolahpour2016strategic}
E.~Nasrolahpour, S.~J. Kazempour, H.~Zareipour, and W.~D. Rosehart, ``Strategic
  sizing of energy storage facilities in electricity markets,'' \emph{IEEE
  Transactions on Sustainable Energy}, vol.~7, no.~4, pp. 1462--1472, 2016.

\bibitem{Xu2017storagestack}
B.~Xu, Y.~Wang, Y.~Dvorkin, R.~Fernández-Blanco, C.~A. Silva-Monroy, J.-P.
  Watson, and D.~S. Kirschen, ``Scalable planning for energy storage in energy
  and reserve markets,'' \emph{IEEE Transactions on Power Systems}, vol.~32,
  no.~6, pp. 4515--4527, 2017.

\bibitem{Huang2019storage}
Q.~Huang, Y.~Xu, and C.~Courcoubetis, ``Financial incentives for joint storage
  planning and operation in energy and regulation markets,'' \emph{IEEE
  Transactions on Power Systems}, vol.~34, no.~5, pp. 3326--3339, 2019.

\bibitem{Qin2019storage}
J.~Qin, S.~Li, K.~Poolla, and P.~Varaiya, ``Distributed storage investment in
  power networks,'' in \emph{2019 ACC}, 2019, pp. 1579--1586.

\bibitem{zhao2019virtual}
D.~Zhao, H.~Wang, J.~Huang, and X.~Lin, ``Virtual energy storage sharing and
  capacity allocation,'' \emph{IEEE transactions on smart grid}, vol.~11,
  no.~2, pp. 1112--1123, 2019.

\bibitem{fundamentals}
D.~S. Kirschen and G.~Strbac, \emph{Fundamentals of power system
  economics}.\hskip 1em plus 0.5em minus 0.4em\relax John Wiley \& Sons, 2018.

\bibitem{leslie2020designing}
G.~W. Leslie, D.~I. Stern, A.~Shanker, and M.~T. Hogan, ``Designing electricity
  markets for high penetrations of zero or low marginal cost intermittent
  energy sources,'' \emph{The Electricity Journal}, vol.~33, no.~9, p. 106847,
  2020.

\bibitem{calidata}
\BIBentryALTinterwordspacing
``California iso open-access data,'' accessed on 2021.9.10. [Online].
  Available: \url{http://oasis.caiso.com/mrioasis}
\BIBentrySTDinterwordspacing

\end{thebibliography}

\end{document}